\documentclass{caosp306}

\usepackage{graphicx}
\usepackage{natbib}
\articleNo{000}
\pubyear{2019}
\volume{00}
\volnumber{0}
\firstpage{1}
\received{July 30, 2019}
\accepted{-- --, 2019}

\def\BibTeX{{\rm B\kern-.05em{\sc i\kern-.025em b}\kern-.08em
             T\kern-.1667em\lower.7ex\hbox{E}\kern-.125emX}}

\begin{document}

\hauthor{S.\,Dodonov, A.\,Grokhovskaya}

\title{The density maps of the HS47.5-22 field }

\author{
        S.\,N.\,Dodonov \inst{1} 
      \and 
        A.\,Grokhovskaya \inst{1}   
       }

\institute{
           Special Astrophysical Observatory\\
           Russia
          }

\date{July 30, 2019}
\maketitle

\begin{abstract}
We study the reconstruction of overdensity maps of galaxies as function of redshift in the range $0 < \mathrm z < 0.8$ using data from 1-m Schmidt Telescope of Byurakan Astrophysical Observatory (Armenia) in 16 medium band ($\sim 250 \mathrm \AA$) and four broad band (u,g,r,i) filters. The data used in this work homogeneously cover $2.39$ sq. deg with accurate photometric redshiftss, down to $\mathrm R < 23$ mag (AB). We reconstructed the density contrast maps for the whole galaxy sample of the HS 47.5-22 ROSAT field in narrow slices for full range of redshifts. We select groups and clusters of galaxies with adaptive kernel based on density peaks which are larger than two times the mean density. The reconstructed overdensity field of galaxies consists of cluster-like structures outlining void-like regions for full redshift range $0 \leq \mathrm z \leq 0.8$. We detect known galaxy clusters in this field with software specially developed for this project. This gives us a possibility to study how star formation properties and galaxy morphology depend on the environments of the galaxies in this field.

\keywords{methods -- data analysis -- galaxies -- clusters}
\end{abstract}

\section{Introduction}
\label{intr}

Galaxy evolution and physical properties must be in strong correlation with their environment. There are a several dependencies of galaxy physical properties and the environment. The dependence "morphology - density of the environment" was discovered by \cite{Oemler1974} and \cite{Dressler1980}. These authors stated that the early type galaxies are more often located in overdensity areas in the center of groups and clusters , and that the late type galaxies are preferentially found in the periphery of groups and clusters. More recent studies based on 2dFGRS \citep[Two-degree-Field Galaxy Redshift Survey,][]{Madgwick2003} and SDSS \citep[Sloan Digital Sky Survey,][]{Guo2013,Guo2014} have found that this correlation is maintained for the entire range of local densities up to the field galaxies.

For nearby galaxies the dependencies of $\mathrm  {H_\alpha}$ equivalent width, $4000 \; \mathrm\AA$ break (or the ratio between the average flux density in $\mathrm {ergs \; s^{-1} \; cm^{-2} \; Hz^{-1}}$ between 4050 and $4250 \; \mathrm {\AA}$ and that between 3750 and $3950 \;  \mathrm \AA$ \citep{Bruzual1983}) and colors on the environment were found on scales of the order of 1 $ \mathrm {Mpc \; h^{-1}}$ \citep{Kauffmann2004}. In the recent article \cite{Cucciati2010} it was shown that more massive galaxies were formed in the most dense areas earlier than galaxies with a smaller mass and the evolution of less massive galaxies occurs under the influence of complex physical processes determined by their environment.

Galaxies located in areas of high density of groups and clusters of galaxies are formed and evolve differently than galaxies in the low density voids. One of the main methods for determination of galaxies belonging to a large-scale
structure, in the optical and infrared ranges, is density contrast maps based on filtering algorithms \citep{Lopes2004,Koester2007}.

In this paper we describe the survey observations and the data reduction in Section 2. We review filtering algorithm with adaptive kernel, basic statistics and overdensity field reconstruction in Section 3. We discuss our results and make conclusion in Section 4. Throughout this paper, we assume a flat cosmology described with $\Omega_m = 0.3$, $\Omega_\Lambda = 0.7$ and $ \mathrm{H_0 = 70 \; km \; s^{-1}}$.

\section{Observations}
\label{Obs}

\begin{figure}
\centerline{\includegraphics[width=0.85\textwidth,clip=]{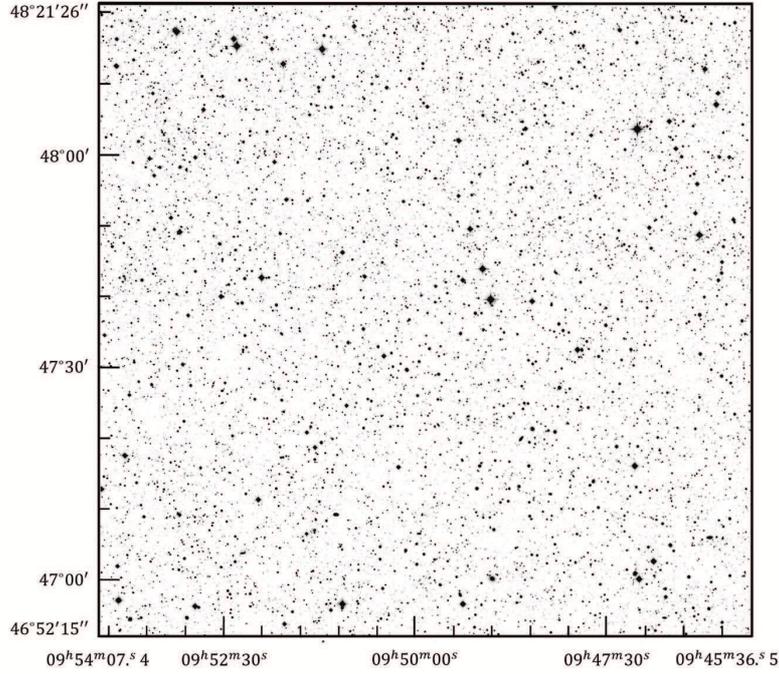}}
\caption{The composite optical image in g, r and i filters of the HS47.5-22 field is about 2.39 sq. degree with center coordinates $09^h50^m00^s \; +47^d35^m00^s$. 574 X-ray sources up to level $\rm 3.5 \cdot 10^{-14} erg\,cm^{-2}\,s^{-1} $ were detected by the ROSAT X-ray satellite \citep{Molthagen1997}.}
\label{field}
\end{figure}

We observed the HS47.5-22 ROSAT field (Fig. \ref{field}) defined by the medium deep ROSAT survey with center coordinates $09^h50^m00^s \; +47^d35^m00^s$ \citep{Molthagen1997}. The survey consists of 48 overlapping ROSAT pointings which were added to produce a final catalog containing 574 X-ray sources with broad band ($0.1-2.4 \; \mathrm{keV}$) count rates between $\rm \sim  3 \times 10^{-3} \;cts\, s^{-1}$ and $\rm \sim 0.2 \; cts\, s^{-1}$, in a field of view of $\sim\; 2.3 \;\mathrm{sq. \;deg.}$. \cite{Molthagen1997} use an X-ray error circle of $2  \sigma + 10$ arcsec in radius, with $\sigma $ taken from ROSAT observations. $\sigma $ is positioning error box for ROSAT objects.

Observations of the field HS47.5-22 were carry out with 1-m Schmidt Telescope of the Byurakan Observatory (Armenia) during several sets in February, March, April and November of 2017 and in February and November of 2018 years. Telescope field of view with $4k \times 4k$ CCD was $58 \times 58$ arcmin, scale $0.868 \; \mathrm{arcsec \; {pixel}^{-1}}$. To get data for near all pointings of HS47.5-22 ROSAT field we observe four positions with 10 arcmin overlaps. From these observations we create mosaic with total area of 2.386 sq. degree \citep{Dodonov2017}. 

\begin{table}[ht]
\caption{1m Schmidt Telescope Filter set. Effective Wavelength, $FWHM$, Limiting Magnitude measured at F/2.} 
\centering 
\begin{tabular}{c c c c} 
\hline\hline 
Filter & $\lambda_{\mathrm{cen}}, \AA$ & $FWHM (\AA)$  & $m_{\mathrm{lim},5\sigma}$\\ [0.5ex] 
\hline 
u\_SDSS & 3578 & 338 & 24.23 \\ 
g\_SDSS & 4797 & 860 & 25.22 \\
r\_SDSS & 6227 & 770 & 24.97 \\
i\_SDSS & 7624 & 857 & 24.15 \\
MB\_400 & 3978 & 250 & 24.37 \\
MB\_425 & 4246 & 250 & 24.31 \\
MB\_450 & 4492 & 250 & 24.20 \\
MB\_475 & 4745 & 250 & 24.31 \\
MB\_500 & 4978 & 250 & 24.30 \\
MB\_525 & 5234 & 250 & 24.37 \\
MB\_550 & 5496 & 250 & 23.86 \\
MB\_575 & 5746 & 250 & 24.29 \\
MB\_600 & 5959 & 250 & 23.89 \\
MB\_625 & 6234 & 250 & 23.51 \\
MB\_650 & 6499 & 250 & 23.41 \\
MB\_675 & 6745 & 250 & 23.78 \\
MB\_700 & 7002 & 250 & 23.47 \\
MB\_725 & 7253 & 250 & 23.20 \\
MB\_750 & 7519 & 250 & 23.07 \\
MB\_775 & 7758 & 250 & 22.97 \\[1ex] 
\hline 
\end{tabular}
\label{tabl1} 
\end{table}

\begin{figure}
\centerline{\includegraphics[width=0.75\textwidth,clip=]{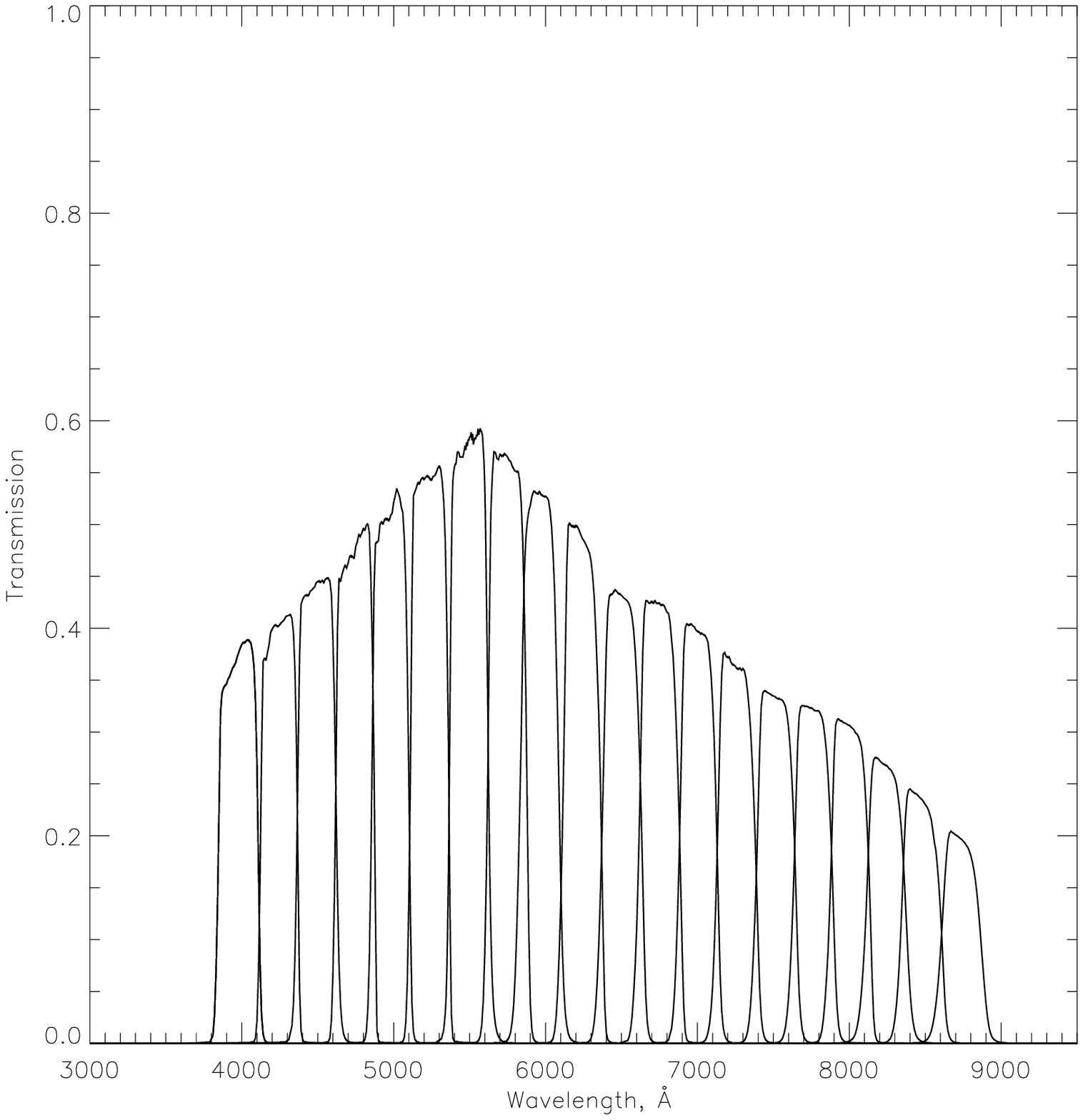}}
\caption{1m Schmidt Telescope Filter set. Filter transmission measured at F/2. CCD spectral response included.}
\label{filter}
\end{figure}

Each position was observed in four broad band filters ($u$, $g$, $r$ and $i\; SDSS$) and in 16 medium band ($FWHM=250 \; \AA$, FWHM is full width at half maximum) filters with homogeneous covering optical spectral range from $4000 \; \AA $ till $ 8000 \; \AA$ (see Tab. \ref{tabl1} and Fig. \ref{filter}). In each filter we obtained from 12 to 25 individual images with 5-6 arcmin shifts between each one. Total exposure time in filters were varied from 60 min to 120 min depending from the spectral sensitivity of the CCD. Long term objects variability controlled by the observations in $r-SDSS$ filter in each set of observations. 

Photometry of the objects obtained using SExtractor \citep{Bertin1996} in dual image mode. Base image were created from the sum of deep ($\sim 25^m$) images obtained in $g$, $r$ and $i$ filters. Before the photometry all images were convolved to common seeing quality and transformed to common coordinates system. We get photometry in $1.45 \times FWHM$ apertures and correct received fluxes for light loss using light curve obtained from bright stars, and using MAG\_AUTO SExtractor photometry we receive Kron-like fluxes for all objects in the field. Photometric calibration was developed using spectral and photometrical data from SDSS survey for the objects detected in the field. By using field objects as a standard stars within each exposure, we were independent from photometrical conditions for imaging.

Galaxies sample is extracted from the full catalog of objects (near 100000 objects) using following criteria:
\begin{itemize}
\item Objects brighter then $R_{AB}=23$ mag (AB magnitude in R filter)
\item Extended index $\textless \; 0.8$ \citep{Bertin1996} for the objects with $R_{AB}<21$ mag, extended index $\textless \; 0.9$ for the objects with $21 \; \mathrm{mag} \; <R_{AB}<22$ mag and extended index $\textless \; 0.96$ for the objects with $R_{AB}<23$ mag  
\item Index of contamination $\leq 2$ \citep{Bertin1996}
\end{itemize}

The sample of objects which follow the first two criteria includes 40194 objects and after applying the third one - we have 32637 objects with clean photometry. Due to the contamination we lose 8.12 \% of the objects. 

\begin{figure}
\centerline{\includegraphics[width=0.6\textwidth,clip=]{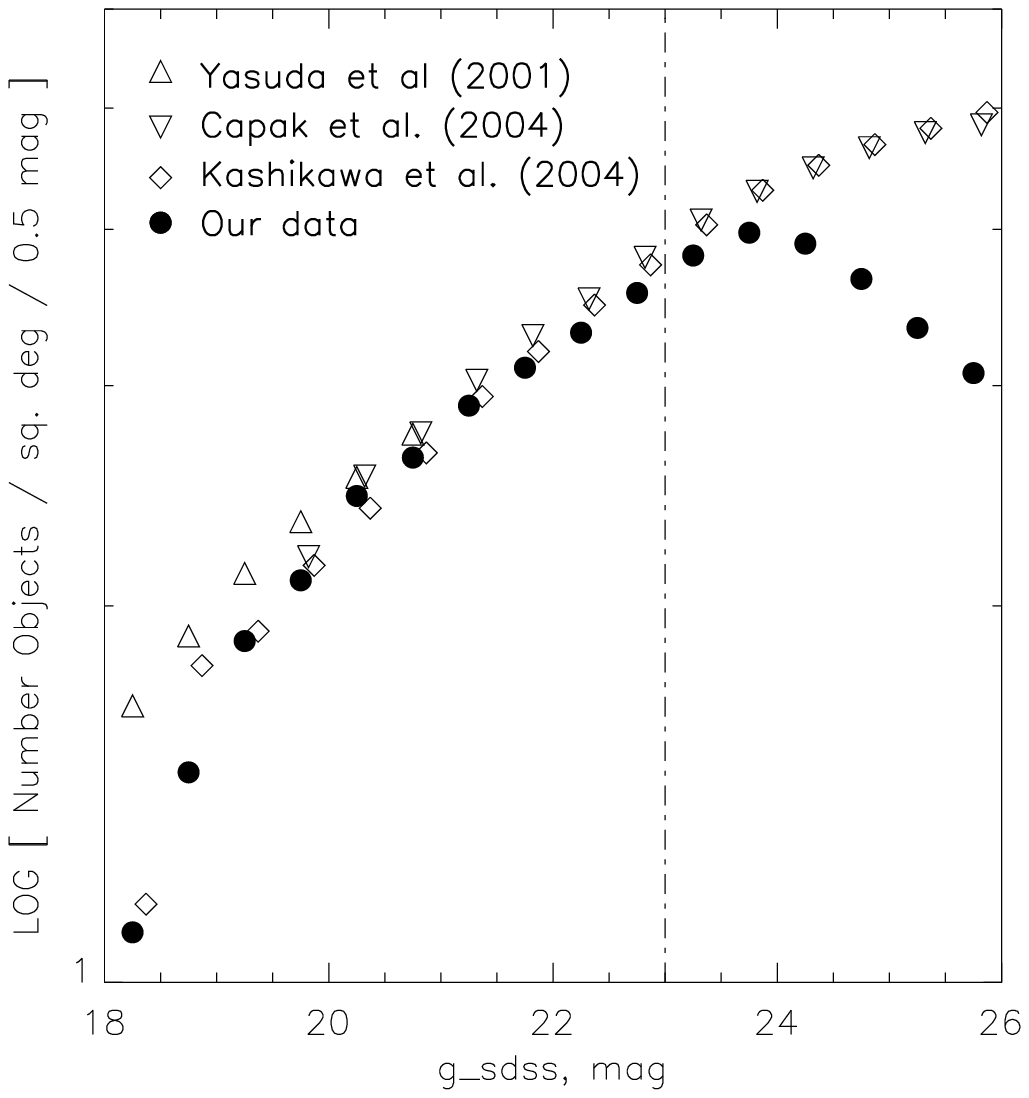}}
\centerline{\includegraphics[width=0.6\textwidth,clip=]{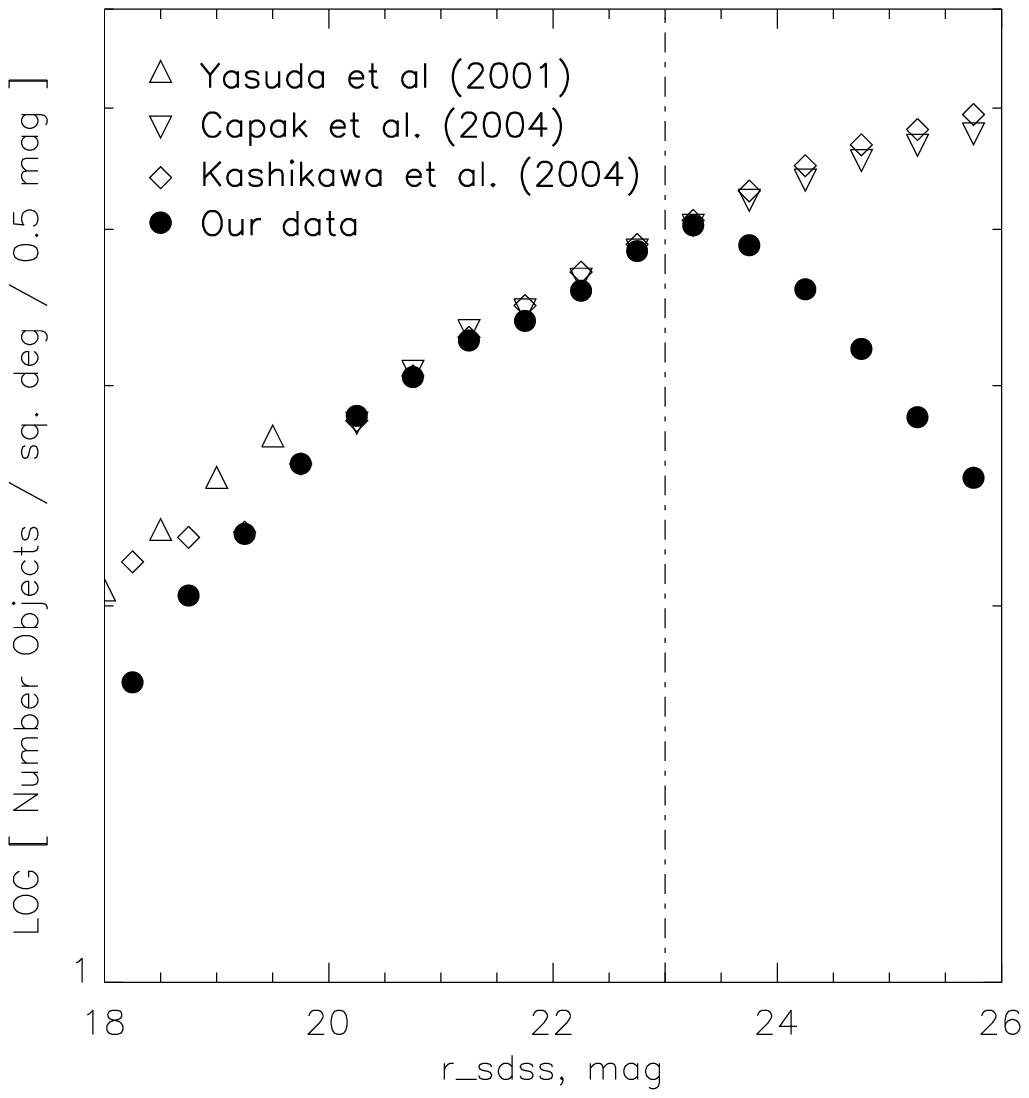}}
\centerline{\includegraphics[width=0.6\textwidth,clip=]{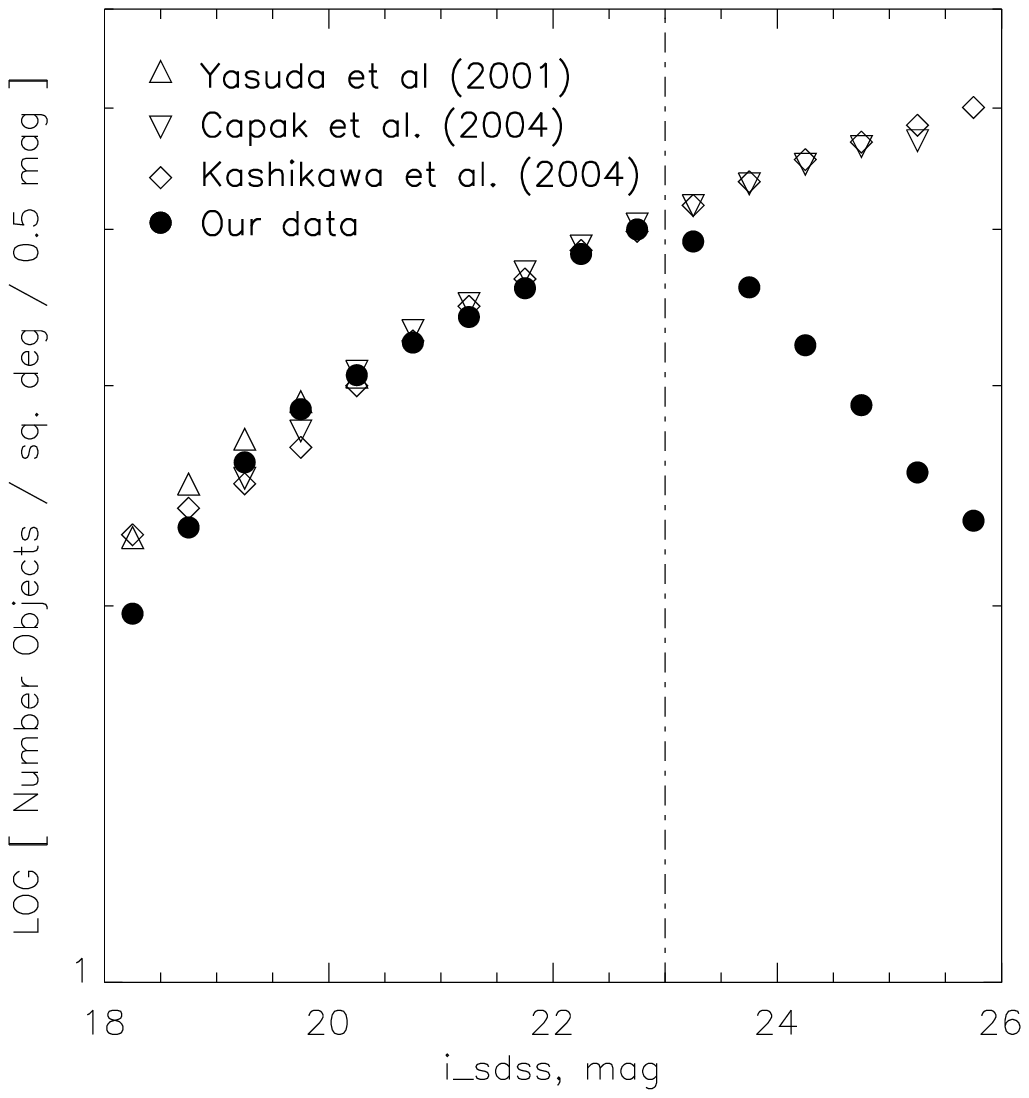}}
\caption{Galaxy sample completeness in $g$, $r$ and $i \; SDSS$ filters.}
\label{counts}
\end{figure}

We check sample completeness using comparison of galaxies number-counts in $g$, $r$ and $i \; SDSS$ filters from our sample with already published data in Fig. \ref{counts}. The galaxies sample completeness up to $R_{AB}=23$ mag with no color selection effects in all optical range. 

\begin{figure}
\centerline{\includegraphics[width=1.2\textwidth,clip=]{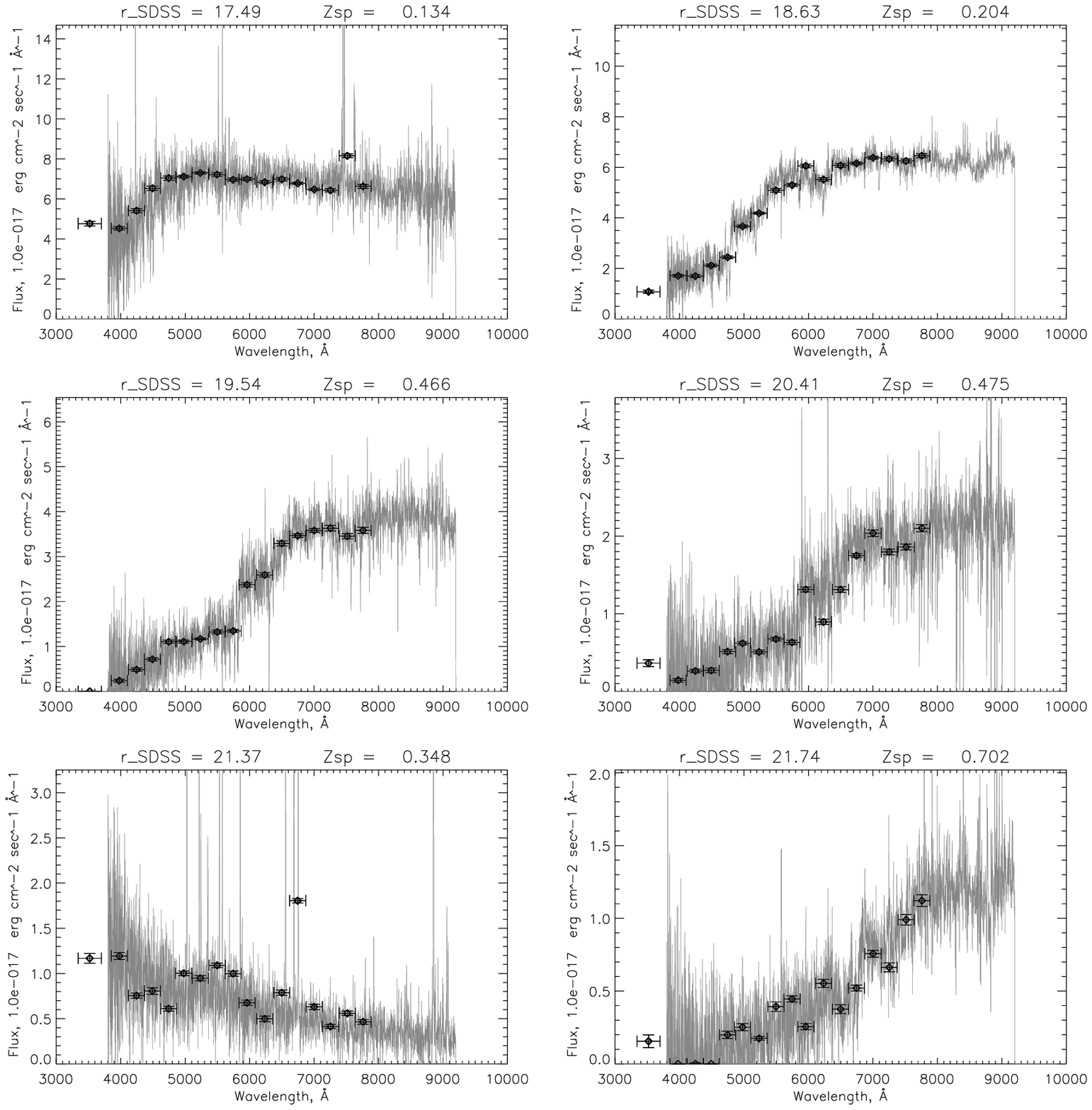}}
\caption{Comparison of the Spectral Energy Distribution (SED) obtained from the photometric data with 1-meter Schmidt telescope (points with bars) with the SDSS spectra (solid gray line). Points in the SED are scaled to 3 arcsec to match the SDSS fiber diameter. Horizontal bars corresponds to filter width and vertical bars to 1$\sigma$ flux error.}
\label{SDSSvsSED}
\end{figure}

Photometric measurements from 17 filters ($u-SDSS$ plus 16 medium band filters) provide low resolution spectra Fig. \ref{SDSSvsSED} for each object which are analyzed by a statistical technique for classification and redshift estimation based on spectral template matching. We used for these galaxies spectra templates library from \cite{Dodonov2008}, and a set of programs ZEBRA  \citep[Zurich's Extragalactic Bayesian Redshift Analyzer,][] {Feldmann2006}.

\begin{figure}
\centerline{\includegraphics[width=0.75\textwidth,clip=]{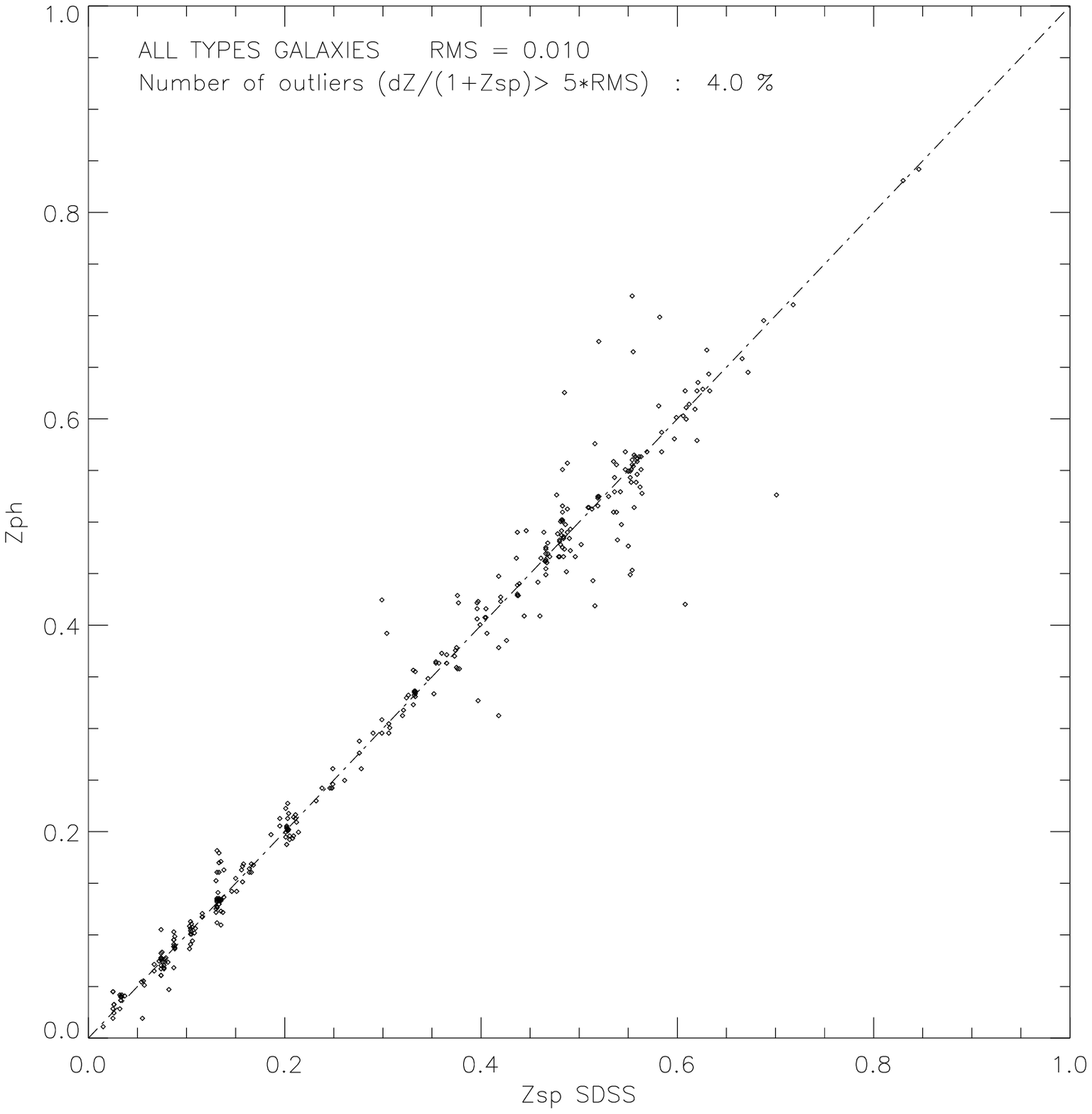}}
\centerline{\includegraphics[width=0.75\textwidth,clip=]{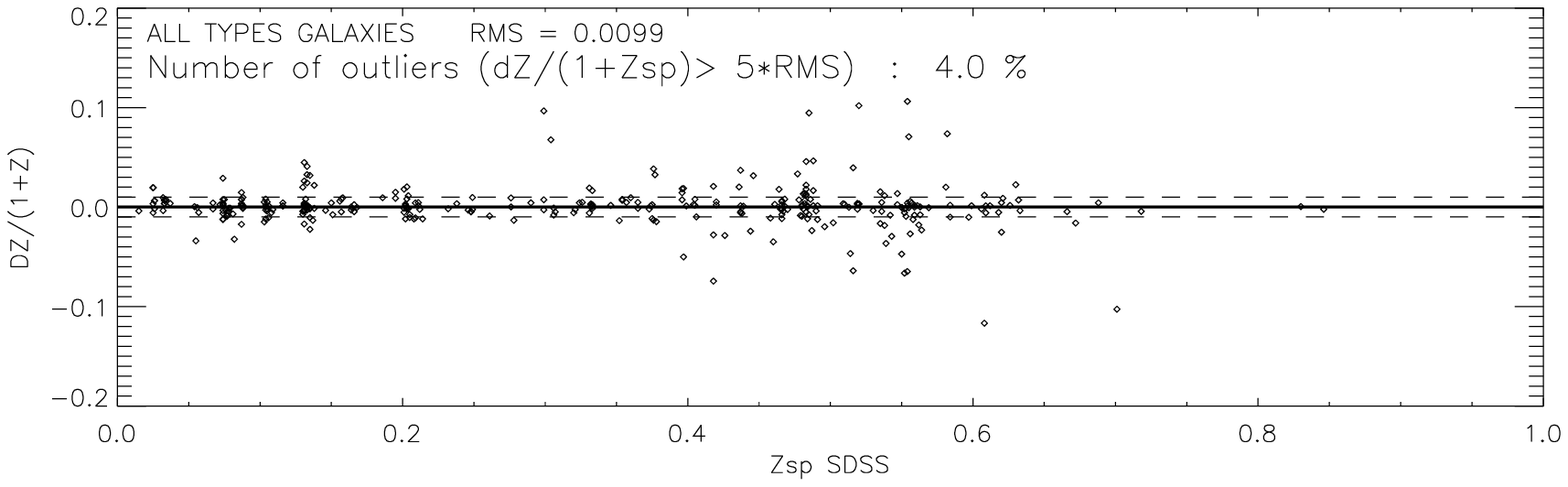}}
\caption{Comparison between photometric redshifts $z_{ph}$ obtained with ZEBRA in Maximum Likelihood mode with SDSS spectroscopic redshifts $z_{sp}$ along with error distribution $\Delta z/(1+z)$ for all galaxies with known spectroscopic redshifts}
\label{redshifts}
\end{figure}

The obtained redshift accuracy $\sigma_z < 0.01$ and the fraction of catastrophic outliers is ($\Delta z/(1+z) > 5.*\sigma_z$)  $\sim 4.0\% $ (in Fig. \ref{redshifts}). Accuracy $\sigma_z$ changes from 0.01 in magnitude range $r-SDSS = 16 \; \mathrm{mag} - 21 \; \mathrm{mag} $ till 0.03 in magnitude range $r-SDSS  = 21\; \mathrm{mag} - 23 \; \mathrm{mag}$.

\section{Data analysis}
\label{Analysis}

\subsection{Filtering algorithm with adaptive kernel}

In this work we use the algorithm with adaptive kernel to reconstruct the density contrast field. In this method we calculate the density value in the vicinity of each galaxy:
$$\delta_i=\frac s {\frac43 \pi R^3},$$
where $R$ is three dimensional distance from the galaxy to its N-th nearest neighbor and $s$ is number of nearest neighbor galaxy. Next we divide the light cone into thin redshift slices with $\Delta z = 0.056 \cdot (1 + z)$ and 25\% overlap from each side of slice. The width of slices is based on $1\sigma = 0.028$ photometric redshift error for all type galaxies. For each slice the mean density is estimated as 
$$\overline \delta = {\frac 1 n}  \sum_{i=1}^n \delta_i,$$
where $n$ is the total number of galaxies in the each slice.

The density contrast $\sigma_i$ for each galaxy position is calculated as
$$\sigma_i+1=\frac {(\delta_i-\overline \delta)} {\overline \delta} +1.$$
The candidate to clusters and group of galaxies are detected as density peaks which are larger than two times the average density and the candidate to voids are detected as density cavity smaller than ten times the average density in each redshift slice.

\subsection{Basic statistics}

For testing filtering algorithm with adaptive kernel we use galaxy mock-catalog MICECAT v2 \citep{Carretero2017}. It is a fake or simulated galaxy catalogue which resembles a genuine galaxy redshift survey catalogue, but is built from a cosmological simulation. Mock catalog is useful for cluster-determination algorithms because it has information about dark halo id. Thus we can estimate statistical parameters of detecting clusters sample over mock sample. Obviously, it is impossible to obtain a perfect match between structures of the model catalog and the recovered groups and clusters. Moreover completeness and purity of reconstructed structures tend to be mutually exclusive. In the process of determining the elements of a large-scale structure, it is possible to obtain the following cases: over-merging (several real groups combined to one reconstructed group), fragmentation (one real group divided to a several reconstructed groups), spurious group (reconstructed group with no one real group matching), undetected group (real group is missing in reconstructed catalog).

Basic statistic term are: I) the “completeness”, which is a measure of the fraction of real groups with N or more members that are successfully recovered in the reconstructed group catalogue, II) the purity, which is a measure of the fraction of reconstructed groups with N or more members that belong to real groups, III) "galaxy success rate" which is the fraction of galaxies belonging to real groups of special richness that have ended up in any reconstructed group, and IV) "interloper fraction" is the fraction of galaxies belonging to reconstructed groups of special richness which are field galaxies (\citet{Knobel2010}, see also \citet{Grokh2019}).

\begin{figure}[h]
\begin{minipage}[h]{0.49\linewidth}
\center{\includegraphics[width=1.0\linewidth]{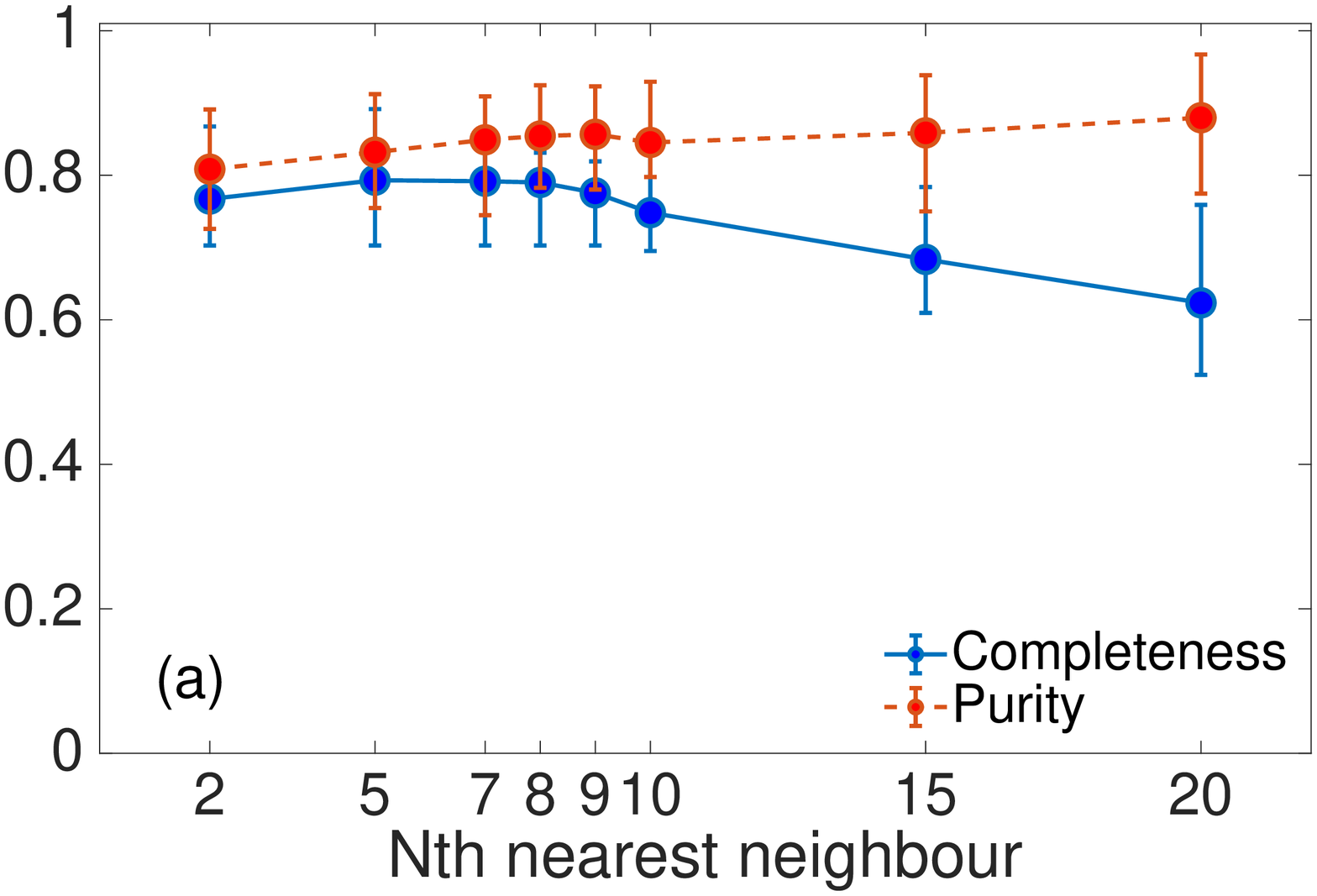} \\}
\end{minipage}
\hfill
\begin{minipage}[h]{0.49\linewidth}
\center{\includegraphics[width=1.0\linewidth]{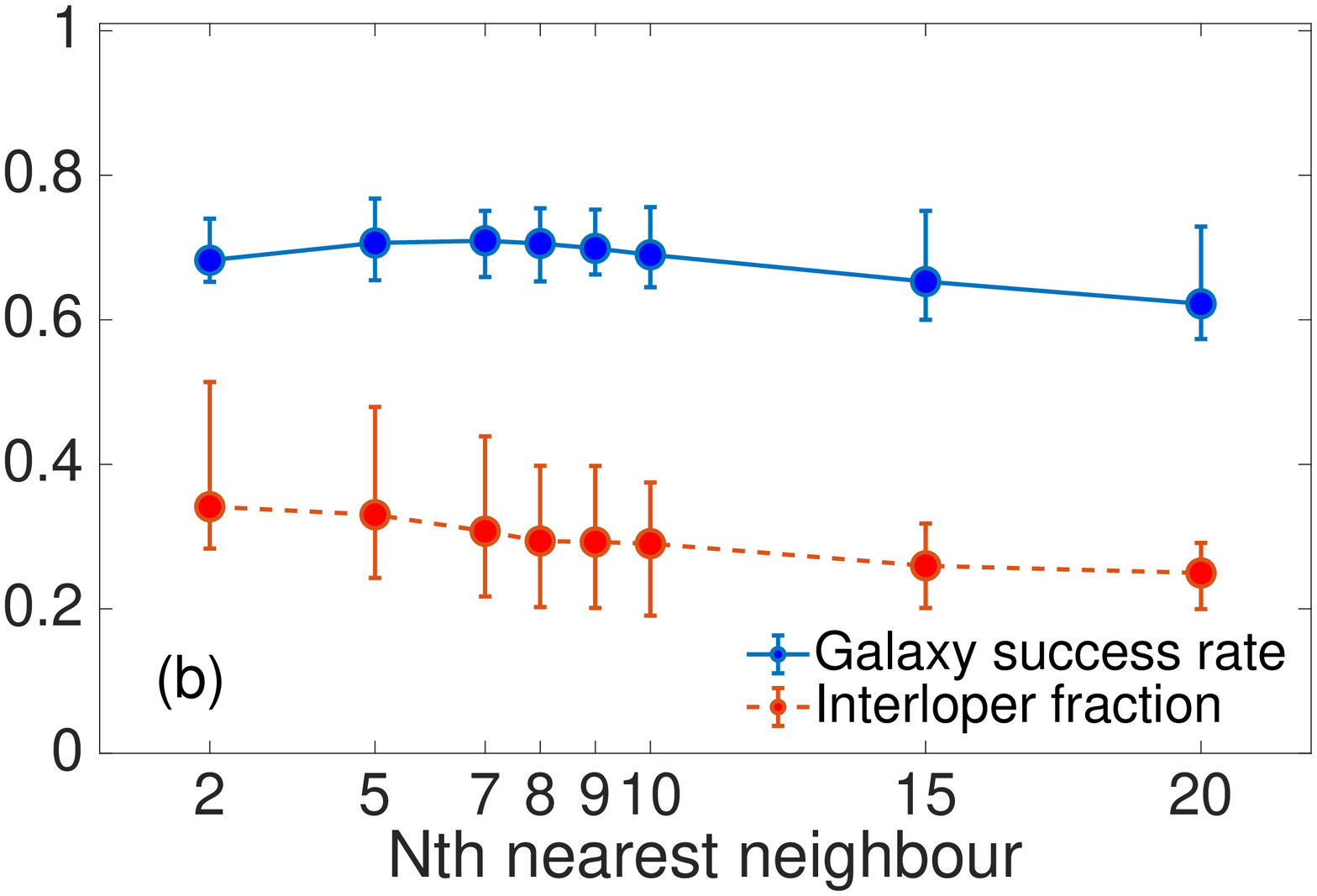} \\}
\end{minipage}
\caption{Basic static parameters of the sample of groups, clusters and galaxies obtained by filtering algorithm with adaptive kernel. On the left the figures show completeness (solid line) and purity (dashed line) of the sample, on the right - the percentage of successful detection of galaxies from clusters of the model catalog (solid line) and the percentage of galaxies defined automatically as galaxies of clusters, but being field galaxies in the model catalog (dotted line). The bars correspond to the standard deviation for 10 samples of galaxies of the MICE simulation.}
\label{ris:image1}
\end{figure}

We use 10 mock samples with the same physical properties like observation data (2 sq. degree field, z $\leq$ 0.8 and $R_{AB} \; \leq \; 23$ mag) for estimation of all basic statistical parameters for detected clusters by filtering algorithm with adaptive kernel (Fig. \ref{stat}). For this method we can variate size of aperture as number of nearest neighbor for calculation and compare statistics for choosing best size of aperture. The 8 nearest neighbours is optimal variant of completeness and purity.

\subsection{The HS47.5-22 overdensity field}

We reconstructed 3D overdensity maps for observational data of the HS 47.5-22 field. We obtained a broad range of reliably reconstructed local overdensities of the sample of galaxies by using filtering algorithm with variable aperture. The size of aperture was defined by the distance to the 8th nearest neighbour from basic statistics. Based on the tests on the mock catalogue, 8 is the optimal number of neighbours which can be used to reliably reconstruct density at all redshifts. An increase in the size of the adaptive aperture (e.g. 10, 15 or 20) leads to a smoothing of the peaks on the contrast maps of the density of galaxies and the underdetermination of large-scale structures. Decrease in the size of the adaptive aperture (e.g. 2, 5 or 7) leads to decrease of the purity of the reconstructed cluster sample.

\begin{figure}
\centerline{\includegraphics[width=0.9\textwidth,clip=]{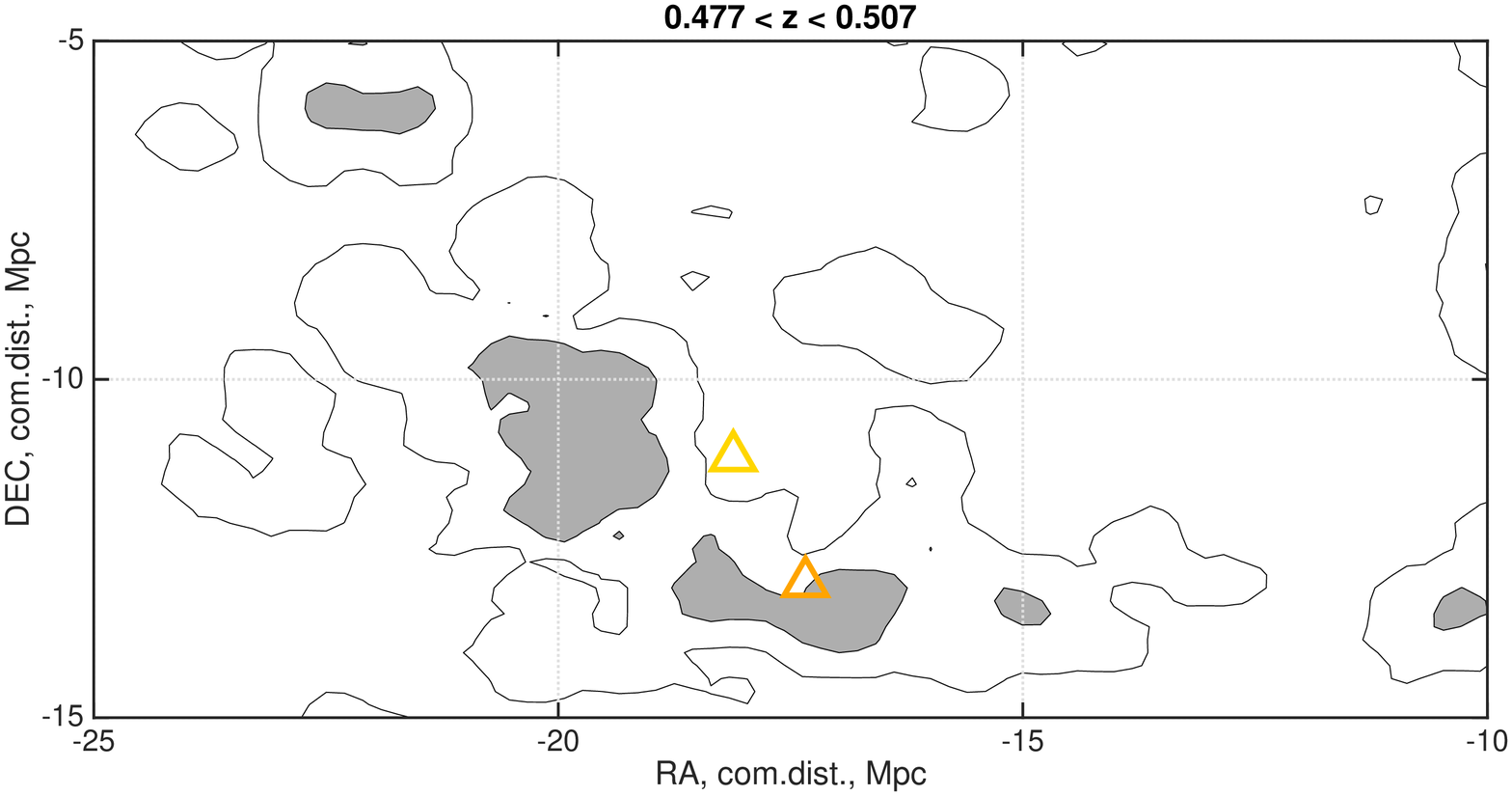}}
\centerline{\includegraphics[width=0.9\textwidth,clip=]{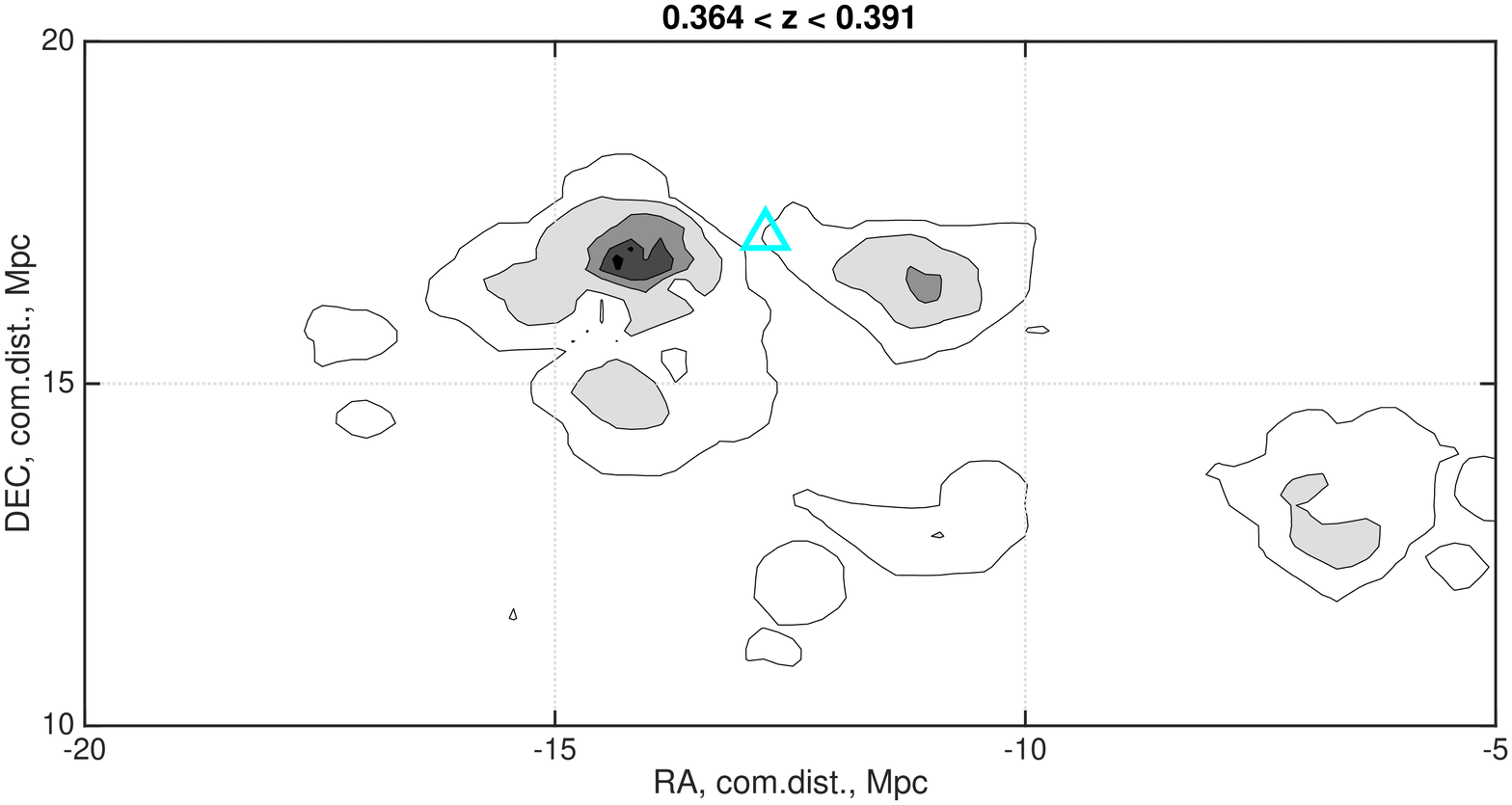}}
\centerline{\includegraphics[width=0.9\textwidth,clip=]{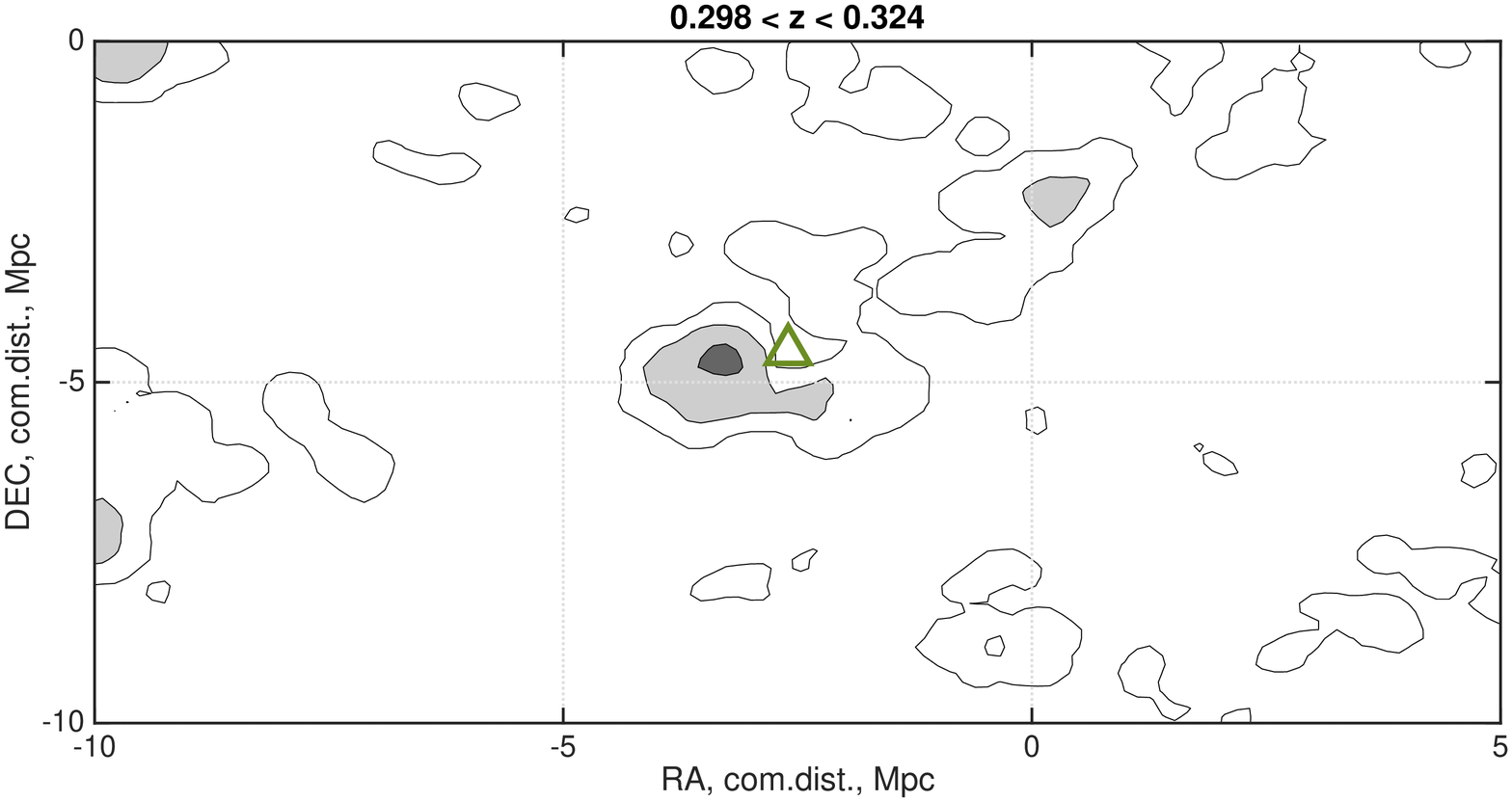}}
\centerline{\includegraphics[width=0.9\textwidth,clip=]{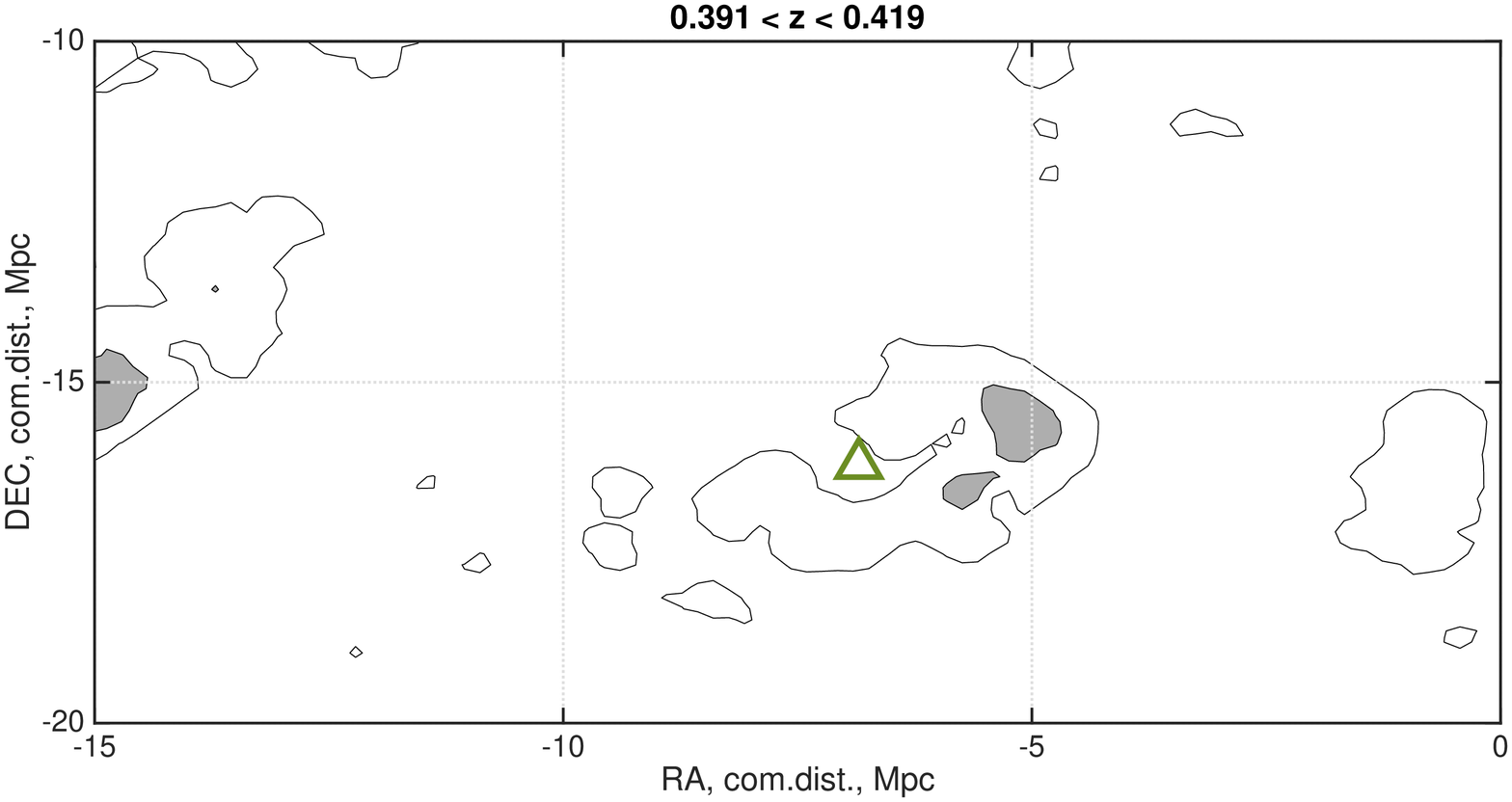}}
\caption{Overdensity maps (grey surfaces are isosurfaces with a density contrast two times above the mean density mean density) of narrow slices HS 47.5-22 field with WHL clusters (black triangles, from top to bottom): 1. WHL J094645.7 + 471107 and WHL J094637.9 +471440 2. WHL J094657.3 +481526 3. WHL J094913.0 +472248 4. WHL J095027.9 +481435. Axis are shown in comoving distance (x-axis is distance from the survey centre in right ascension, y-axis is distance from the survey centre in declination)}
\label{WHL_clusters}
\end{figure}

Figure \ref{WHL_clusters} shows isosurfaces for a density contrast which are larger than two times the average density (grey surfaces) and clusters with spectral redshifts from WHL (by the name of the authors of the catalog Wen, Han and Liu) catalog (black triangles) which was completed by \cite{Wen2015} from spectroscopic and photometric data of the SDSS. 

\section{Conclusion}
\label{Conclusion}
We used observational data from 1-m Schmidt Telescope of the Byurakan Observatory (Armenia) to reconstruct the overdensity maps in the HS47.5-22 ROSAT field. We explored the photometric properties of the sample of 36447 galaxies in the field and obtained spectral types and photometric redshifts for all objects. An accuracy of redshift allows you to determine whether a galaxy belongs to a cluster or group.

We applied the filtering algorithm with adaptive aperture for the reconstruction. This algorithm have been tested on the MICECAT v2 mock catalog. We estimated basic statistical parameters (completeness, purity, galaxy success rate and interloper fraction) for reconstructed clusters from mock catalog. Evaluation of the main statistical parameters allowed us to choose the optimal aperture size for further work with observational data. 

We obtain a broad band of large scale structures up to $z=0.8$ for observational data. Also we found the most part of clusters from WHL cluster catalog with spectroscopic redshift in our field.

Obtained results allow us to begin a study of the connection between star formation rate in galaxies and their position in the large scale distribution. 

\acknowledgements

This work has made use of CosmoHub. CosmoHub has been developed by the Port d'Informacio Cientifica (PIC), maintained through a collaboration of the Institut de Fisica d'Altes Energies (IFAE) and the Centro de Investigaciones Energeticas, Medioambientales y Tecnologicas (CIEMAT), and was partially funded by the "Plan Estatal de Investigacion Cientifica y Tecnica y de Innovacion" program of the Spanish government.  

The methodological part of the work was supported by the Russian Science Foundation under grant no. 17-12-01335 (observation, development of methods for processing and calibrating data). 

The data analysis and the study of the large-scale structure were performed as part of the government contract of the SAO RAS approved by the Ministry of Science and Higher Education of the Russian Federation.

\bibliography{dodonov_grokh}

\end{document}